# High-Mobility Two-Dimensional Electron Gases at Oxide Interfaces: Origins and Opportunities

Y. Z. Chen (陈允忠)[a),*], N. Pryds[a)], J. R. Sun (孙继荣)[b)], B. G. Shen (沈保根)[b)], and S. Linderoth[a)]

[a)]*Department of Energy Conversion and Storage, Technical University of Denmark, Risø Campus, Roskilde 4000, Denmark*

[b)]*Beijing National Laboratory for Condensed Matter Physics and Institute of Physics, Chinese Academy of Sciences, Beijing 100190, China*

The discovery of two-dimensional electron gas (2DEG) at well-defined interfaces between insulating complex oxides provides the opportunity for a new generation of all-oxide electronics. Particularly, the 2DEG at the interface between two perovskite insulators represented by the formula of $ABO_3$, such as $LaAlO_3$ and $SrTiO_3$, has attracted significant attention. In recent years, progresses have been made to decipher the puzzle of the origin of interface conduction, to design new types of oxide interfaces, and to improve the interfacial carrier mobility significantly. These achievements open the door to explore fundamental as well as applied physics of complex oxides. Here, we review our recent experimental work on metallic and insulating interfaces controlled by interfacial redox reactions in $SrTiO_3$-based heterostructures. Due to the presence of oxygen-vacancies at the $SrTiO_3$ surface, metallic conduction can be created at room temperature in perovskite-type interfaces when the overlayer oxide $ABO_3$ involves Al, Ti, Zr, or Hf elements at the $B$-sites. Furthermore, relying on interface-stabilized oxygen vacancies, we have created a new type of 2DEG at the heterointerface between $SrTiO_3$ and a spinel γ-$Al_2O_3$ epitaxial film with compatible oxygen ions sublattices. The spinel/perovskite oxide 2DEG exhibits an electron mobility exceeding 100,000 $cm^2V^{-1}s^{-1}$, more than one order of magnitude higher than those of hitherto investigated perovskite-type interfaces. Our findings pave the way for design of high-mobility all-oxide electronic devices and open a route towards studies of mesoscopic physics with complex oxides.

---

[*] Corresponding author. Email: yunc@dtu.dk

## 1. Introduction

The realization of high-mobility 2DEGs in epitaxially grown heterostructures made of traditional semiconductors is at the heart of present electronics, which has led to a wealth of new physical phenomena as well as new electronic and photonic devices over the past few decades, starting form p-n junctions to field-effect-transistors and more recently to quantum devices, just to name a few. Currently, widespread interest has arisen in designing heterostructures not from traditional semiconductors but from complex oxides. Different from those in semiconductors, electrons in complex oxides with partially occupied *d*-orbitals interact strongly with one another and with the lattice. This has given rise to a variety of extraordinary electronic properties, such as high-temperature superconductivity, colossal magnetoresistance, ferromagnetism, ferroelectricity, and multiferroicity. Therefore, the high-mobility 2DEGs at atomically engineered complex oxide interfaces are not only expected to provide a wealth of opportunities to study mesoscopic physics with strongly correlated electrons confined in nanostructures, but also show promise for all-oxide devices with functionalities much richer than those found in conventional semiconductor devices.[1,2]

As silicon is the basis of conventional electronics, strontium titanate ($SrTiO_3$, STO), a wide band gap insulator (band gap of 3.2 eV), has been suggested as the foundation of the emerging field of oxide electronics. Particularly, the accessibility of regular single-$TiO_2$-terminated STO substrates[3, 4] triggered the possibility to fabricate complex oxide heterostructures with atomic-scale-flat interfaces through controlled film growth techniques, such as pulsed laser deposition (PLD)[5,6] and molecular beam epitaxy (MBE)[7,8], where the film growth process can be in-situ monitored by reflection high energy electron diffraction (RHEED). In 2004, by the epitaxial growth of another band gap insulator $LaAlO_3$ (LAO, band gap of 5.6 eV) on $TiO_2$-terminated STO substrate, a metallic state was found at the interface between the two insulators (LAO/STO).[9] Since then, many other perovskite-type heterostructures are found to hold similar interfacial 2DEGs as long as STO is involved as the substrate, for instance $LaTiO_3$/STO,[10] $GdTiO_3$/STO,[11] $LaGaO_3$/STO[12] and so on. Moreover, a series of intriguing physical properties, such as, two-dimensional superconductivity,[13] magnetism[14], and field-induced insulator-metal transition[15-17] have also been observed at these oxide interfaces.

Due to the fascinating physics and potential device applications, extensive research activity, both experimental and theoretical,[18-22] has been focused on this 2DEG at perovskite-type oxide interfaces, especially for the LAO/STO system. However, many important issues remain open. One primary question is: What are the origin and the underlying mechanism for the 2DEG at the LAO/STO oxide interface? Firstly, the conduction of the LAO/STO heterostructure is found to depend strongly on the growth parameters, particularly the partial pressure of oxygen, $P_{O2}$, during film growth. At a $P_{O2}$ of $1\times10^{-4}$ Pa and a typical deposition temperature of 800 ºC, a quite high carrier density in the order of $4\times10^{16}$ cm$^{-2}$, as well as a relative high mobility in the order of 1-$2\times10^{4}$ cm$^{2}$V$^{-1}$s$^{-1}$ at 2 K, is routinely obtained.[9, 23, 24] However, samples survive under oxygen annealing or those grown under higher oxygen pressure exhibit lowered carrier density of about 1-2 $\times10^{13}$ cm$^{-2}$ and a typical mobility of 1000 cm$^{2}$V$^{-1}$s$^{-1}$ at 2 K.[9-25] The emerging consensus is that the high carrier density of $4\times10^{16}$ cm$^{-2}$ is due to the electrons released by oxygen vacancies in the bulk STO substrate and they have a 3D character, while the lower carrier density of $10^{13}$ cm$^{-2}$ may be an intrinsic feature of the interface. However, the origin of these intrinsic interface charge carriers is strongly debated. The prevalent interpretation relies on the polar discontinuity at the interface.[9, 26] The STO terminated planes, $(Ti^{4+}O^{2-}_{2})^{0}$ and $(Sr^{2+}O^{2-})^{0}$, are formally neutral in the simple ionic limit, while the LAO planes, $(La^{3+}O^{2-})^{+}$ and $(Al^{3+}O^{2-}_{2})^{-}$, have an alternating $\pm e$ charge. The charge discontinuity between $(Ti^{4+}O^{2-}_{2})^{0}/(La^{3+}O^{2-})^{+}$ or $(Sr^{2+}O^{2-})^{0}/(Al^{3+}O^{2-}_{2})^{-}$ could create a built-in potential in the growing LAO film. The accumulation of this potential to a large enough value at a certain film thickness will result in an injection of carriers from the film surface into the interface, thus forming a 2DEG. In principle, the polar catastrophe can be avoided if 0.5 of an electron per unit cell accumulates in the interfacial TiO$_2$ plane. This amounts to an electron carrier density of $3.4 \times 10^{14}$ cm$^{-2}$, approximately one order of magnitude higher than the sheet density commonly obtained. It is also notable that, the experimentally detected electric field accumulated in the LAO film is much lower than the predicted value.[27] On the other hand, the "polar catastrophe" model is based on the assumption that the interface is chemically abrupt. Nevertheless, ions transfer across the interface and formation of defects have been identified to play important roles on the transport properties of LAO/STO interfaces. For example, there is strong evidence of La-interdiffusion into STO, which can act as electron donor.[28, 29] Additionally, in the growth of oxide heterostructures, it is the match between anion (oxygen) sublattice rather than cation sublattice that plays a dominant role. However, the possible oxygen ions redistribution across the oxide-oxide interface, which is usually associated with strong electron redistribution on neighboring cations, is poorly understood.[30-33] For instance, although most of bulk-like oxygen vacancies in STO-base heterostructures can be removed by suitable annealing, it remains unclear whether some content of oxygen vacancies can be stabilized at the interface by a space charge effect. These interface-stabilized oxygen vacancies, if survive unexpectedly the annealing in high oxygen pressure, would play a nontrivial role on the interface conduction since oxygen vacancies act as one type of the most effective electron donors in STO.

Another striking issue is that the carrier mobilities at oxide interfaces are significantly lower than those obtained in bulk STO materials. Specifically, the typical mobility for the intensively investigated LAO/STO interface is approximately 1000 $cm^2V^{-1}s^{-1}$ at 2 K. This value is still much lower than those for three-dimensional oxygen-deficient STO single crystals[34] and La-doped STO epitaxial films[8], amounting to $1.3 \times 10^4$ $cm^2V^{-1}s^{-1}$ and $3.2 \times 10^4$ $cm^2V^{-1}s^{-1}$, respectively. Though there are reports of enhanced mobility at LAO/STO interfaces by defect engineering,[35] it remains unclear whether the carrier mobilities are limited by sample quality or whether they can be enhanced by the use of different material combinations.

The mechanism that plays the dominant role in the formation of 2DEGs at polar oxide interfaces has not been unambiguously determined. The ways toward enhanced carrier mobility, the far more sensitive test of material cleanliness than any structural or chemical characterization methods, also remain elusive. In semiconductor 2DEGs, the achievement of high mobilities has been accompanied by the observation of new phenomena---first the integer Quantum Hall effect (QHE),[37] and later, in even cleaner samples, the fractional quantum Hall effect (FQHE).[38] These discoveries were the subjects of the 1985 and 1998 Nobel Prizes in physics, respectively. In the progress of oxide electronics, similar phenomena have already been observed in ZnO-based oxide heterostructures.[39,40] It therefore becomes extremely interesting to investigate these charming physical phenomena at complex oxide interfaces with strongly-coupled *d*-orbital electrons, which may open new avenues for condensed matter physics. In this review, we will discuss our recent experimental work on an alternative approach to create oxide 2DEGs by interfacial redox reactions.[33] Replying on interface confined redox reactions, an oxide 2DEG with electron mobility exceeding 100 000 $cm^2V^{-1}s^{-1}$ at 2 K has been realized.[36]

**2. Metallic and insulating interfaces controlled by interfacial redox reactions**

To date, complex oxide 2DEGs have been almost exclusively focused on polar interfaces consisting of two perovskite-type oxides, particularly those heterostructures based on $SrTiO_3$. Interestingly, it remains unclear why each conducting interface has to involve STO, and whether 2DEGs can be obtained when the capping film is made into amorphous rather than crystalline, i.e. when the interface polarity is strongly depressed.

Motivated by the simple question of "why each 2DEG involves STO?", we found that chemical redox reactions at the oxide interface can provide an alternative approach to create 2DEGs.[33] Regardless of interface polarity, when depositing certain insulating oxides with higher oxygen affinity on the STO single crystalline substrate, the overlayer could absorb oxygen from STO. This can alter the wideband insulator of stoichiometric STO into a semiconductor, metal, superconductor, or even a magnet, depending on the concentration of oxygen vacancies. For instance, such interfacial chemical redox reaction results in a remarkable metallic interface between the amorphous LAO film and the crystalline STO substrate, *a*-LAO/STO, the properties of which turn out to be quite similar to its all-crystalline counterpart.[41-43]

The *a*-LAO/STO was grown by PLD at room temperature. Compared to the conventional high temperature film growth procedure, the room temperature deposition has the advantage that it rules outs any annealing-induced oxygen vacancies in STO substrates. Moreover, the possibility of La-interdiffusion, as frequently observed in the crystalline LAO/STO sample grown at high temperatures, should be also strongly suppressed during the deposition at room-temperature.

Figure 1(a) shows a high angle annular dark field (HAADF) scanning transmission electron microscopy (STEM) image for the *a*-LAO/STO sample.[41] A well-defined amorphous-crystalline interface is confirmed. Further electron energy loss spectroscopy (EELS) measurements suggest that the interface is atomically sharp without significant intermixing.[41] The surface morphology of the *a*-LAO/STO, as illustrated in Fig. 1(b), shows a flat terrace-like surface with step-height of 0.4 nm, similar to that of the STO substrate. As shown in Fig. 1(c)-(e), the *a*-LAO/STO shows a metallic behaviour in the investigated temperature range of 2-300 K despite that both *a*-LAO and STO are highly insulating themselves. The sheet carrier density, $n_s$, which dominates by electrons, is nearly constant in the temperature range of 100-300 K with a value of $1.2 \times 10^{14}$ cm$^{-2}$ [Fig. 1(d)]. At $T<100$ K, $n_s$ decreases with decreasing temperature and reaches $\sim 6\times 10^{13}$ cm$^{-2}$ at 2 K. Such a partial carrier freezing-out effect upon cooling probably indicates the presence of trapped electrons with a typical activation energy of 10-20 meV as reported for amorphous CaHfO$_3$/STO interfaces.[44] The electron mobility, $\mu_s$, increases upon cooling and saturates around $\sim$300 cm$^2$ V$^{-1}$s$^{-1}$ at 2 K [Fig. 1(e)]. Notably, the carrier density and the carrier mobility for the $a$-LAO/STO heterostructures is of the same order of magnitude as those reported for the crystalline LAO/STO deposited at high temperatures as also shown in Fig.1 (c)-(e).

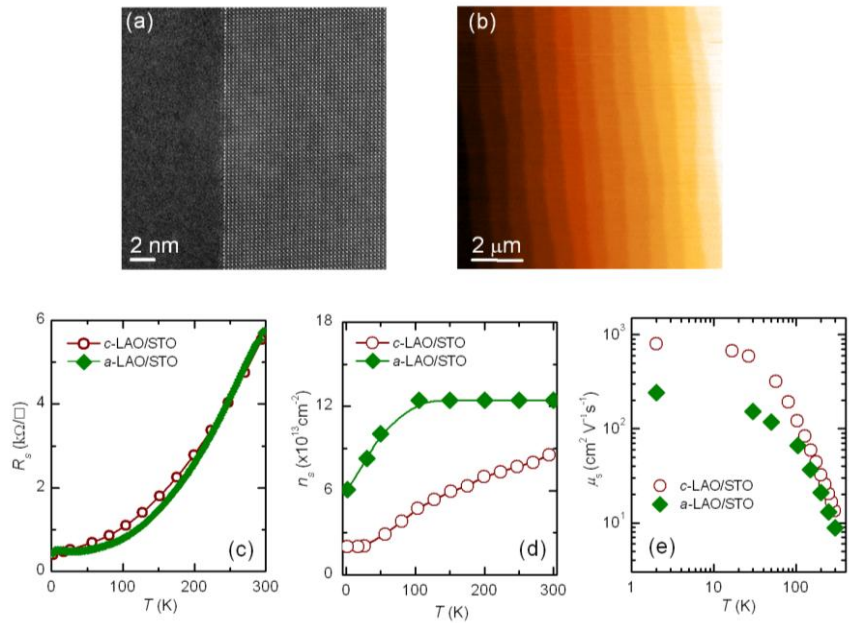

**Fig.1 Metallic conduction at the interface between amorphous LaAlO$_3$ and crystalline SrTiO$_3$ ($a$-LAO/STO).** (a) HAADF-STEM image of the $a$-LAO/STO interface;[41] (b) Smooth terrace-like surface of $a$-LAO/STO measured by atomic force microscopy; (c)-(e) Sheet resistance, carrier density and mobility as a function of temperature, respectively, for the $a$-LAO/STO interface deposited at 0.01 Pa. For comparison, the typical results for crystalline LAO/STO are also demonstrated.

Similar to crystalline LAO/STO interfaces, the $a$-LAO/STO heterostructure exhibits two characteristic properties. Firstly, the interfacial conductivity of the $a$-LAO/STO heterostructures exhibits strong dependence on the $P_{O2}$ during film growth. As shown in Fig. 2, for samples with film thickness around 30 nm, the heterointerfaces grown at $P_{O2}$

>1 Pa are all highly insulating, with $R_s$ equal to that of the bare STO substrates ($R_s > 10^9$ ohms per square, measurement limit). Upon decreasing the pressure below 1 Pa, the *a*-LAO/STO heterointerface turns to conductive dramatically. Note that similar behaviour has also been observed in the crystalline LAO/STO interface, except that the high temperature film growth results in the reduction of the whole STO substrate at the low pressure range of $10^{-4}$ Pa.[14,24] It is also notable that the deposition of amorphous $La_{1-x}Sr_xMnO_3$ (x=1/8)(*a*-LSMO) films on STO substrates does not result in any interfacial conductivity, even though the films were deposited in high vacuum ($P_{O2} \leq 10^{-4}$ Pa).[33] Secondly, an interfacial metal-insulator transition that depends on the film thickness is also observed in *a*-LAO/STO heterostructures. For films deposited at $P_{O2} \approx 1 \times 10^{-4}$ Pa, the *a*-LAO/STO heterointerface is insulating at $t < 1.8$ nm, while these heterointerfaces abruptly become metallic upon further increase of the film thickness. Note that this critical film thickness for the occurrence of interface conduction is nearly equal to that of *c*-LAO/STO ($t = 4$ uc, ~1.6 nm).[15] The threshold film thickness of $t \approx 1.6$ nm in *c*-LAO/STO has been regarded as the minimum thickness requirement for the occurrence of charge transfer to alleviate the "polar catastrophe" [15,26]. However, the "polar catastrophe" is absent in the amorphous-crystalline samples investigated here since there is no long-range translational symmetry in the amorphous overlayers. As will be discussed later, the critical thickness in the *a*-LAO/STO heterostructures may indicate the existence of a threshold concentration of oxygen vacancies where the localized conduction region evolves to be overlapped. This viewpoint is also consistent with the fact that the critical thickness for the occurrence of metallic conductivity of the *a*-LAO/STO heterointerface, is increased upon increasing $P_{O2}$.[42,43]

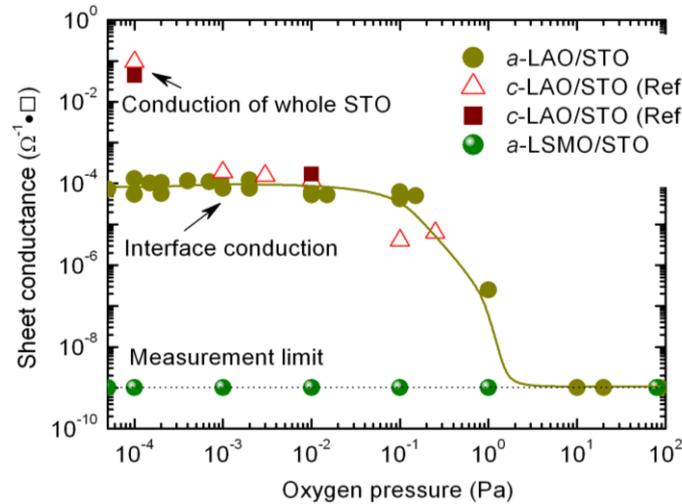

**Fig.2** The dependence of sheet conductance of *a*-LAO/STO on the oxygen pressure of film growth (*T*= 300 K, film thickness of about 30 nm).[42] Results reported for the crystalline LAO/STO samples[14,24] are also shown for comparison.

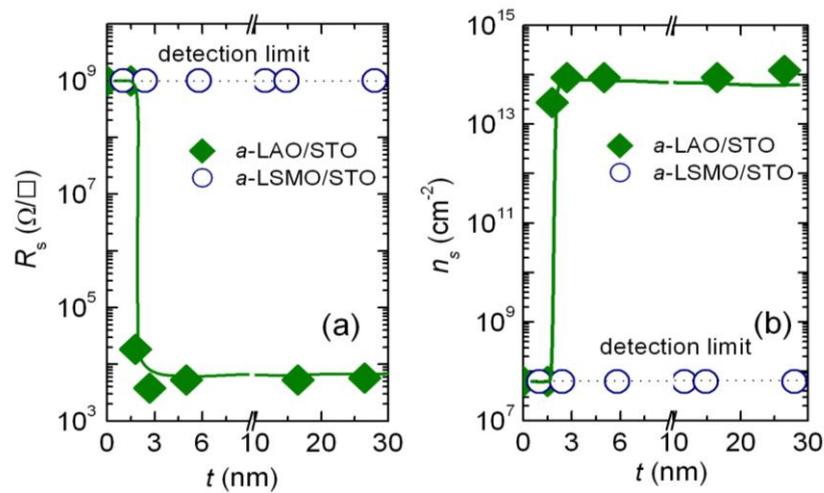

**Fig.3** The critical dependence of sheet conductance of *a*-LAO/STO on film thickness.

*In-situ* X-ray photoelectron spectroscopy (XPS) measurements give direct evidence that the conductivity of the *a*-LAO/STO heterointerface results from the formation of oxygen vacancies on the STO side.[33] Figures 4 (a) and (b) show the Ti $2p_{3/2}$ spectra for different film thicknesses of *a*-LAO/STO grown at $P_{O2} \approx 1 \times 10^{-4}$ Pa. As expected, no clear $Ti^{3+}$ signal in the $2p_{3/2}$ core-level spectra could be detected in the bare STO substrate. However, finite amount of $Ti^{3+}$ is already present in the insulating samples, even at $t = 0.4$ nm. This suggests the formation of defects in STO, either by intermixing or oxygen vacancies. The prominent feature of the XPS result is that the amount of $Ti^{3+}$ increases with increasing film thickness, as shown in Fig. 4(b). This indicates that the oxygen vacancies evolves upon film deposition and thus rules out intermixing. Such conclusion is further confirmed by the fact that annealing $a$-LAO/STO in high oxygen pressure decreased the $Ti^{3+}$ signal to almost zero [Fig. 4(b)] and removed the conductivity completely.

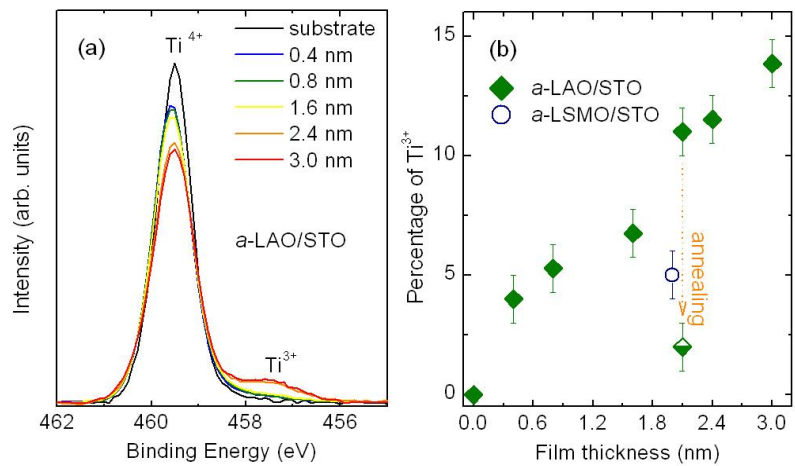

**Fig.4** (a) The Ti $2p_{3/2}$ XPS spectra for different film thicknesses of $a$-LAO/STO grown at $P_{O2}$ ≈1×10$^{-4}$ Pa. The spectra were measured at an emission angle of 80º. (b) The film thickness dependent percentage of $Ti^{3+}$ for $a$-LAO/STO, where $[Ti^{3+}] + [Ti^{4+}] = 100\%$. Annealing the 2.1 nm $a$-LAO/STO sample at 150 ºC for 1.5 hours in 0.6 bar pure $O_2$ reduced the $Ti^{3+}$ signal almost to zero.[33]

It has been generally argued that the oxygen vacancies are created by the bombardment due to the high energy of the arriving species during PLD process at $P_{O2}$ <0.1 Pa.[44,45] However, this viewpoint is hardly compatible with the trend that increasing film thickness enhances the $Ti^{3+}$ content significantly as shown in Fig. 4(a). Furthermore, the bombardment scenario cannot explain the insulating nature of the $a$-LSMO/STO interface grown at $P_{O2}$ ≈1×10$^{-4}$ Pa. Besides the high energy, the plasma species also exhibits high chemical reactivity at $P_{O2}$ <0.1 Pa. Our results strongly suggest that the formation of oxygen vacancies in *a*-LAO/STO hetero-structures result from the outward diffusion of oxygen ions from the STO lattice due to the exposure of the STO surface to the reactive LAO plasma species which show a higher chemical reactivity.

In the scenario of interface redox reaction, the deposited species of the capping film react with the oxygen ions ($O^{2-}$) present in the STO substrate lattice. Interestingly, the defect formation energy of the $TiO_2$-terminated STO surface (5.94 eV), which is considerably smaller than that of the bulk,[46] is comparable to the bond dissociation energy of oxygen molecules (5.11 eV[47]). In this case, besides the oxygen source from the target and the background oxygen, oxygen ions in STO substrates diffuse outward to oxidize the reactive plasma species absorbed on the STO surface, as schematically depicted in Fig. 5 (a). It should be noted that though it has been confirmed that STO substrates can act as oxygen source for the oxide film growth at high temperatures,[30-32] the room temperature outward diffusion of oxygen ions from the STO substrate during the growth of oxide film by PLD has not been recognized previously. On the other hand, similar outward diffusion of oxygen ions from the STO substrate has been observed during the room-temperature growth of reactive metals of Ti and Y films by molecular beam epitaxy under ultrahigh vacuum.[48]

In the case of metal-oxide interface, the redox reaction can be regarded as an electron transfer process from metal atoms to the $Ti^{4+}$ accompanied with the oxygen ion reconstruction. The chemical interactions at the interface between a metal and the $TiO_2$ or STO surface are controlled not only by the thermodynamic stability of the metal oxide, but also by the space charges at the metal/oxide interface, which is determined by the interface electronic configuration, i.e. the relative Fermi level of the metal and that of the $TiO_2$ or STO before contact.[49] An interfacial redox reaction can occur at room temperature when the formation heat of metal oxide, $\Delta H_f^O$, is lower than -250 kJ/(mol O) and the work function of the metals, $\varphi$, is in the range of 3.75 eV$<\varphi<$5.0 eV.[49] For oxide-oxide interfaces, both the electron transfer and the redox reaction across the interface are still poorly understood. Taking $ABO_3$ perovskites into account, where $A$ is an electropositive cation such as an alkali, alkaline-earth or rare-earth ion, the electropositive character of the $A$-site ion minimizes its contribution to electronic states near the Fermi level, $E_F$. Generally, in these oxides the top of the valence band is primarily oxygen 2$p$ non-bonding in character, while the conduction band arises from the $\pi^*$ interaction between the $B$ site transition metal $t_{2g}$ orbitals and oxygen.[50-52] Therefore, electron transfer across the perovskite-type interface is expected to depend primarily on the electronegativity and coordination environment of the transition metal ions on the $B$ site.

**Table 1** Metallic and insulating perovskite-type interfaces based on $SrTiO_3$ (STO)

| System | Interface polarity | $B$ site reactivity | Interface Conduction |
|---|---|---|---|
| $LaAlO_3$/STO[9] | ✓ | ✓ | ✓ |
| $LaTiO_3$/STO[10] | ✓ | ✓ | ✓ |
| $LaGaO_3$/STO[12] | ✓ | ✓ | ✓ |
| $LaCrO_3$/STO[53] | ✓ | ✗ | ✗ |
| $LaMnO_3$/STO[54] | ✓ | ✗ | ✗ |
| Amorphous $LaAlO_3$/STO[33] | ✗ | ✓ | ✓ |
| Amorphous $CaHfO_3$/STO[44] | ✗ | ✓ | ✓ |
| Amorphous STO/STO[33] | ✗ | ✓ | ✓ |
| Amorphous LSMO/STO[33] | ✗ | ✗ | ✗ |

Table 1 summarizes most of the STO-based perovskite oxide interface reported so far. Besides LAO/STO, LaTiO$_3$/STO[10] and LaGaO$_3$/STO[12] are the other two most researched crystalline systems that show metallic interface with STO. Additionally, the crystalline LaCrO$_3$/STO[53] interface is very interesting since it shows interface polarity while exhibits insulating behavior. As for the amorphous STO-based heterointerface, besides the metallic *a*-LAO/STO, amorphous CaHfO$_3$/STO[44] and amorphous STO/STO[33] are the other systems that show metallic interfaces. Additionally, the La$_{1-x}$Sr$_x$MnO$_3$/STO interface is another notable system. The undoped LaMnO$_3$ is polar but it forms an insulating interface with STO.[54] On the other hand, though most crystalline La$_{1-x}$Sr$_x$MnO$_3$ samples are conductive,[55] their amorphous form is insulating and forms an insulating interface with STO.[33] Remarkably, Al, Ti, and Hf, the *B*-site metal for most conductive perovskite-type interfaces, each satisfies well the criterion for metal-oxide interface redox reactions: $\Delta H_f^O <$ -250 kJ/(mol O) and 3.75 eV$< \varphi <$5.0 eV. On the other hand, Mn, Cr and Ga, the main atomic composition of the corresponding La$_{1-x}$Sr$_x$MnO$_3$, LaCrO$_3$ and LaGaO$_3$ plasma, locate on the border region for the occurrence/nonoccurrence of redox reaction of metals on TiO$_2$ surface at room temperature.[49] Though the redox reactions at oxide-oxide interfaces could be more complex than those at metal-oxide interfaces, the lack of interface redox reaction should explain the insulating nature of the LaCrO$_3$/STO and the La$_{1-x}$Sr$_x$MnO$_3$/STO interfaces. As illustrated in Fig.5(b), for the *AB*O$_3$/STO (*A*=La) perovskite-type interfaces, compared to the interface polarity viewpoint, it is surprising to find that the redox reaction scenario can better account for most of the metallic and insulating interfaces reported so far. Moreover, for metallic perovskite interfaces, the *B* site metals locate mainly in the region of $\Delta H_f^O <$ -350 kJ/(mol O) and 3.75 eV$< \varphi <$4.4 eV.

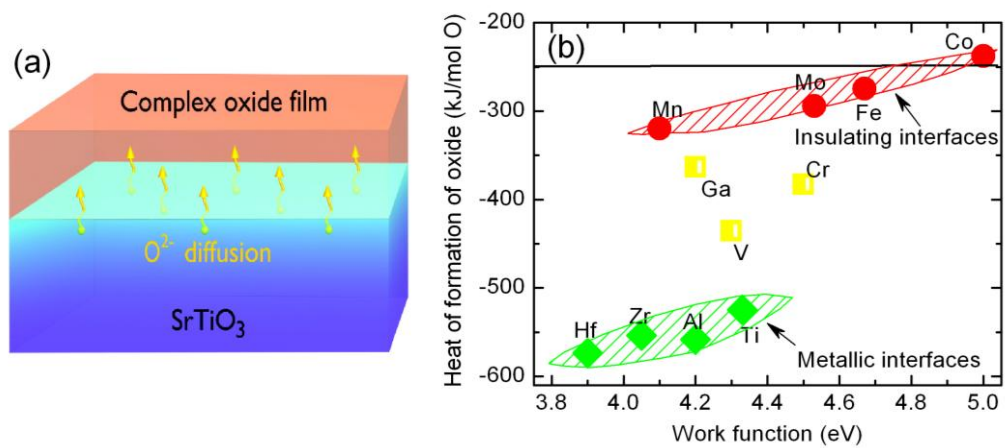

**Fig.5** (a) Sketch of redox reaction at oxide interfaces; (b) Formation heat of metal oxides versus the work function of *B*-site metals for the $ABO_3$/STO (*A*=La) perovskite-type interface. For metallic interfaces, the *B* site metals locate mainly in the region of $\Delta H_f^O < -350$ kJ/(mol O) and 3.75 eV$<\varphi<$4.4 eV (Square and Diamond). The solid line indicates the border of the redox reaction for metals on the STO surface.

Complying with the redox reactions mechanism, other Al-containing oxides have also been used to create an oxide 2DEG with STO as demonstrated by atomic layer deposition or electron beam evaporation.[56,57] It is also notable that the most recent research reveals that Al excess is necessary to assure a metallic interface for *c*-LAO/STO,[58] which further highlights the importance of the interface redox reaction for the formation of 2DEG at LAO/STO interface. Using the concept of interfacial redox reaction, many new opportunities to tune the interface conduction have also been demonstrated. For example, by improving the chemical reactivity of the plasma plume through deposition the LAO film in argon rather than in oxygen, the *a*-LAO/STO conduction can be enhanced.[59] Moreover, the conductivity of *a*-LAO/STO interfaces is also found to be modified by an external electric field provided *in-situ* through a biased truncated cone electrode (-10 V $\leq V_{bias} \leq$ 20 V) during film deposition.[60,61] By modulating the charge balance of the plasma plume, a substantial surface charge is found to transiently builds up during the initial phase of the film growth when the truncated cone electrode is biased. The presence of this surface potential is expected to greatly influence the surface/plasma interaction. More specifically, experimental observations seem consistent with a lowering of the incoming Al-ion flux by the applied bias, which shifts the interfacial conduction of *a*-LAO/STO from metallic over semiconducting to insulating transport mode. This remarkable behaviour further indicates the importance of Al-ion flux on the amount of near-interface oxygen vacancies formed at the STO surface and therefore the interface conduction.

## 3. A high-mobility 2DEG at the spinel/perovskite interface of γ-Al$_2$O$_3$/SrTiO$_3$ (GAO/STO)

In the above section, it has been demonstrated that the redox reaction at oxide interfaces provides a new approach to design metallic and insulating interfaces. However, confining the oxygen vacancies at the oxide interface to realize 2DEGs is still rather challenging, since the high quality epitaxial oxide heterostructures are normally formed at high temperatures, such as at 600 ºC. At such high temperature, oxygen diffuses over many micrometers in minutes because STO is a mixed conductor at high temperatures. This would completely level out any nanometer-scale steps in the oxygen concentration profile in STO-based heterostructures. In this vein, spatial confined redox reactions at oxide interfaces turn out to be a crucial issue to realize high mobility 2DEGs. Fortunately, the previous defect chemistry research of STO implies a positive space charge core at its grain boundary, which is expected to result in the accumulation of electrons in proximity to the interface.[62, 63] This makes the interface-stabilized oxygen vacancies highly practicable, particularly at the atomically engineered oxide interface. Recently, we realized such interface confined oxygen vacancies at an epitaxial spinel-perovskite interface between γ-Al$_2$O$_3$ (GAO) and STO.[36] Remarkably, this new type of oxide 2DEG shows record electron mobilities greater than 140,000 $cm^2V^{-1}s^{-1}$ at 2 K, 100 times higher than those typically obtained in perovskite-type oxide interfaces researched so far.

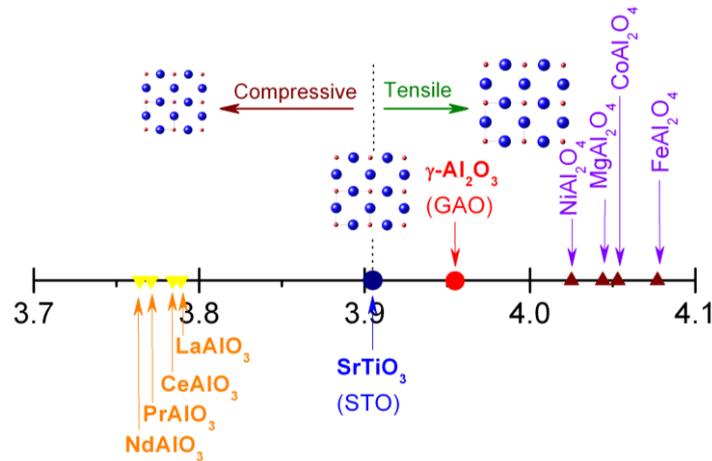

**Fig.6** The lattice parameter diagram for most Al-based oxides which show lattice parameters close to that of STO ($a_{STO}$=3.905 Å).

The most intensively investigated oxide interfaces for 2DEGs are Al-containing perovskites grown on STO, such as the LAO/STO. Since Al satisfies the criteria for redox reactions on the STO surface, any Al-containing oxide has the possibility to exhibit a metallic interface with STO. Fig.6 summarizes the lattice parameter diagram for most Al-based oxides. The relevant perovskite compounds show a smaller lattice parameter than that of STO ($a_{STO}$=0.3905 nm) with lattice mismatches exceeding 3.2% (for LAO/STO). Extending the lattice parameter range to larger than that of STO, one has to rely on the oxides with a spinel structure, the lattice parameter of which is normally larger than 0.4 nm ($a$/2). It is noteworthy that the GAO distinguishes from the other oxides by an excellent lattice match with STO ($a_{GAO}$ =0.7911 nm, lattice mismatch of 1.2%).[64] As shown in Fig. 7 (a) and (b), despite differences in cation sublattices, the oxygen sublattices of perovskite STO match perfectly with the oxygen sublattices of spinel GAO. This makes it possible to grow epitaxially GAO/STO spinel/perovskite heterostructures in a persistent two-dimensional layer-by-layer mode as confirmed by periodic oscillations of the RHEED intensity during film growth [Fig. 7(d)]. Note that for GAO films grown along the (001) direction, one RHEED intensity oscillation corresponds to the growth of one quarter unit cell film, since the GAO unit cell consists of four neutral "AlO$_x$" sub-unit cells with an interlayer distance of about 0.2 nm. Under optimized conditions, persistent sub-unit-cell layer-by-layer growth is observed till more than 200 RHEED oscillations. The obtained GAO/STO heterostructure shows perfect terrace surface with terrace height of 0.4 nm as determined by atomic force microscopy [Fig. 7(e)]. STEM-EELS measurements confirmed a well-defined cubic-on-cubic coherent heterointerface with no obvious dislocations as shown in Fig.7 (f) and (g).

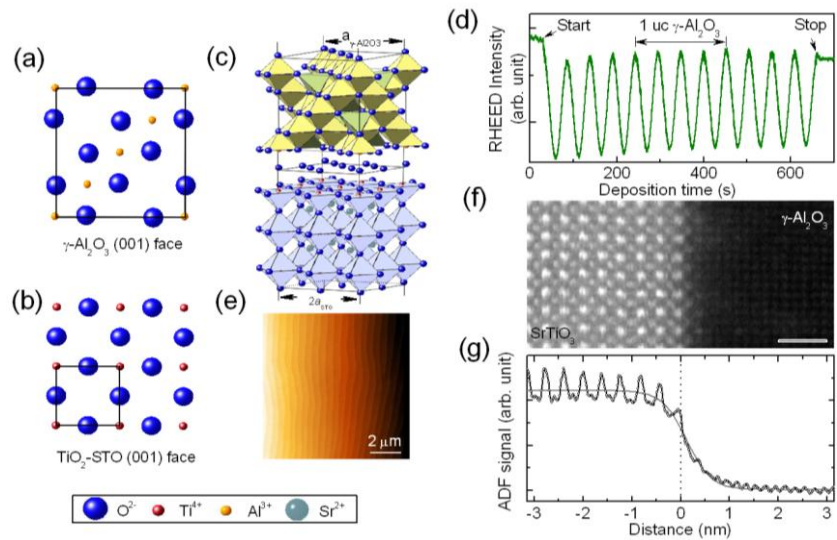

**Fig.7 The epitaxial spinel/perovskite GAO/STO interfaces.** (a) and (b) The compatibility in oxygen sublattices of a GAO surface and the TiO$_2$-terminated STO surface, which forms the backbone to build the spinel/perovskite heterostructure as illustrated in (c). (d) Typical RHEED intensity oscillations for the growth of a 3-uc GAO on STO in a sub-unit-cell layer-by-layer mode.(e) A 10 μm× μm AFM morphology of GAO/STO. (f), High-angle annular dark field (HAADF) STEM image of the epitaxial GAO/STO interface. Scale bar, 1 nm. Sr ions are brightest, followed by Ti. The faintly visible Al elements can be determined by the averaged line profiles across the interface shown in (g). A well developed TiO$_2$-AlO$_x$ heterointerface is defined.[36]

Similar to the LAO/STO interface, the GAO/STO heterostructures can exhibit metallic interface between these two insulators with electrons as the dominant charge carriers. Strikingly, 2DEGs with extremely high Hall electron mobilities are obtained at this spinel/perovskite interface when the GAO film is grown at an oxygen background pressure of 0.01 Pa and a growth temperature of 600 ºC. As shown in Fig. 8, the interfacial conduction depends critically on the thickness, $t$, of the GAO film. The interface conduction occurs only when $t \geq 2$ uc. By carefully controlling the film growth down to a sub-unit-cell level, a great decrease in $R_s$ of approximately four orders in magnitude upon cooling is observed at $t$=2.5 uc [Fig. 8(a)]. Furthermore, the low-field Hall measurements give rise to an impressive $\mu_{Hall}$ of approximately $1.4 \times 10^5$ cm$^2$V$^{-1}$s$^{-1}$ and a $n_s$ of $3.7 \times 10^{14}$ cm$^{-2}$ at 2 K for this highly conductive sample [Fig. 8 (b) and (c)]. Note that the carrier density of the GAO/STO varies dramatically in the range of $10^{13}$-$10^{15}$ cm$^{-2}$ and high mobilities of $\mu_{Hall} \geq 10^4$ cm$^2$V$^{-1}$s$^{-1}$ at $T$=2 K are only detected in the thickness range of 2 uc $\leq t <$ 3 uc. Further increasing $t$ deteriorates the electron mobility to less than 1000 cm$^2$V$^{-1}$s$^{-1}$, probably due to the significant outward diffusion of the Ti-cations across the interface as observed by electron energy loss spectroscopy (EELS).[36]

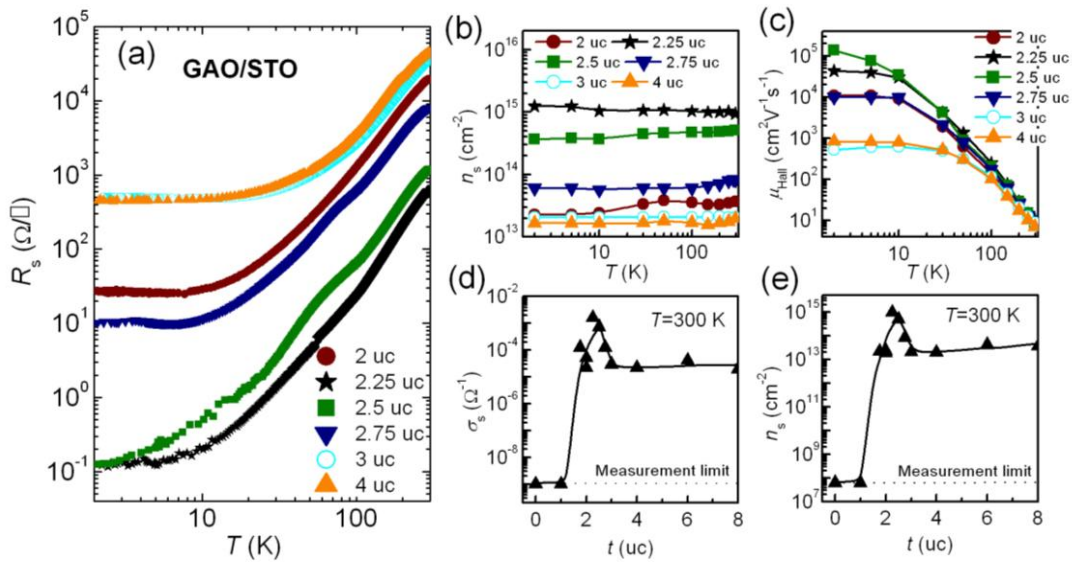

**Fig.8 Thickness-dependent transport properties of the GAO/STO interface.** (a)-(c), Temperature dependence of sheet resistance, $R_s$, carrier density, $n_s$, and low-field electron Hall mobility, $\mu_{Hall}$, for the interface conduction at different film thicknesses. (d) and (e), Thickness dependence of the sheet conductance, $\sigma_s$, and $n_s$ measured at 300 K. High-mobility 2DEGs are obtained in the thickness range of 2 uc ≤ $t$ < 3 uc.[36]

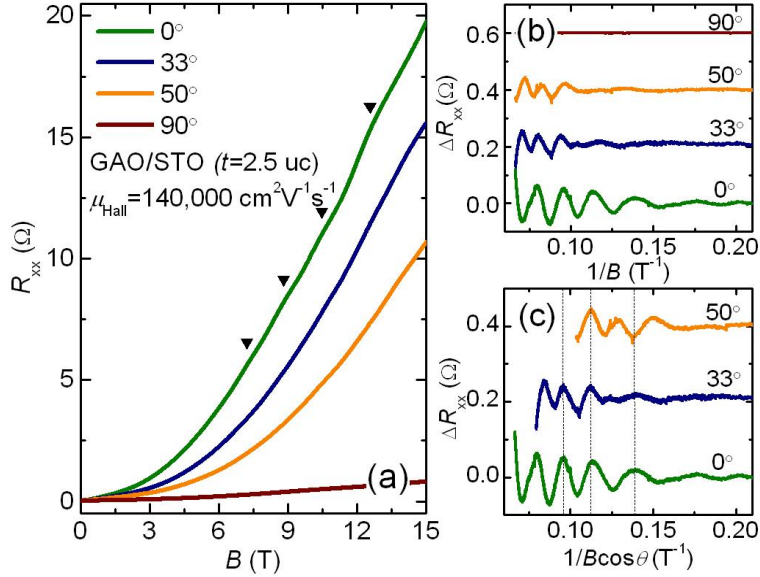

Fig. 9 **Two-dimensional quantum oscillations of the conduction at GAO/STO interfaces.** (a) Longitudinal resistance, $R_{xx}$, as a function of magnetic field with visible SdH oscillations (arrowheads) under different tilt angle, $\theta$, at 0.3 K for the $t$=2.5 uc sample. (b) and (c), Amplitude of the SdH oscillations, $\Delta R_{xx}$, under different $\theta$ versus the reciprocal total magnetic field and the reciprocal perpendicular magnetic field component, respectively. The SdH oscillations depend mainly on the reciprocal perpendicular magnetic field component, particularly in the $\theta$ angle of 0º-33º, which suggests a two-dimensional conduction nature of the GAO/STO interface.[36]

The high mobility 2DEG at the GAO/STO interface exhibits unambiguous angle-dependent Shubnikov-de Haas (SdH) quantum oscillations, which are superimposed on a huge background of positive magnetoresistance [Fig. 9(a)]. After subtracting the magnetoresistance background, the SdH oscillations become apparent as shown in Fig. 9 (b). The extrema positions show a cosine dependence on the angle $\theta$ between the magnetic field and the surface normal [Fig. 9(c)]. This reveals the two-dimensional nature of the electron gas formed at GAO/STO interfaces. Besides, the absence of oscillations at $\theta$ =90° further confirms that the spatial width of the 2DEG is smaller than at least the cyclotron radius at 15 T, the typical value of which is below 10 nm for GAO/STO heterostructures. Moreover, the angular dependence of the SdH oscillations measured at high magnetic fields suggests a multiple-subband contribution to charge transport. For instance, an extra feature is observed at $\theta=50°$ with $B\cos\theta =7.2$ T, which may result from a $\pi$ shift of the oscillations due to a spin-split band. Such a phase shift has been observed in the high-mobility 2DEG of GaN/AlGaN interfaces when the Zeeman energy (depending on the total $B$) and the cyclotron energy (depending on the perpendicular component of $B$) are equal.[65] Importantly, the low-field dependence of the SdH oscillations reveals the typical behavior due to a single band.[36] More quantitative analysis of the temperature dependent SdH oscillations lead to a carrier effective mass of $m^* =(1.22\pm0.03)$ $m_e$ ($m_e$ is the bare electron mass), consistent with those reported for LAO/STO heterostructures.[25,66] Furthermore, the total scattering time $\tau$ for the GAO/STO interface is determined to be $\tau =4.96\times10^{-12}$ s, corresponding to a quantum mobility $\mu_{SdH}=e\tau/m^*$ of $7.2 \times10^3$ cm$^2$V$^{-1}$s$^{-1}$. Such an unprecedented high $\mu_{SdH}$ in our GAO/STO 2DEGs is more than 1 order of magnitude higher than those observed in the perovskite-type LAO/STO[25,66] and GaTiO$_3$/STO[67] heterostructures, which are typically below 300 cm$^2$V$^{-1}$s$^{-1}$. Note that the difference between $\mu_{Hall}$ and $\mu_{SdH}$ could come from a different scattering time (*i.e.* the transport scattering time and the total scattering time, respectively). SdH oscillations also represent a direct measurement of the area of the Fermi surface. For GAO/STO, the typical Fermi cross-section corresponds to a SdH frequency *of F*=14.7 T. By taking a single valley and the spin degeneracy $g_s$ =2, this gives an $n_{2D}= 7.1\times10^{11}$ cm$^{-2}$ under assumption of a circular section of the Fermi surface. As a consequence, the sheet carrier density deduced from the SdH oscillations show significant discrepancy from that obtained from the Hall measurements. This is probably due to the existence of low mobility 2D or 3D subband electrons which do not satisfy the conditions to exhibit SdH oscillations ($\hbar\omega_c \geq k_B T$ and $\omega_c \tau \geq 1$) but contribute nevertheless to the Hall signal. For $n_{2D}= 7.1\times10^{11}$ cm$^{-2}$, the corresponding Fermi wavelength $\lambda_F = 2\pi/k_F = \sqrt{2\pi/n}$ is about 30 nm. The mean free path $l_{mfp}$ of these electrons with a $\tau =4.96\times10^{-12}$ s and $m^*=1.22$ $m_e$ is approximately 100 nm, which is approximately 10 times larger than the $l_{mfp}$ for LAO/STO interfaces.[1]

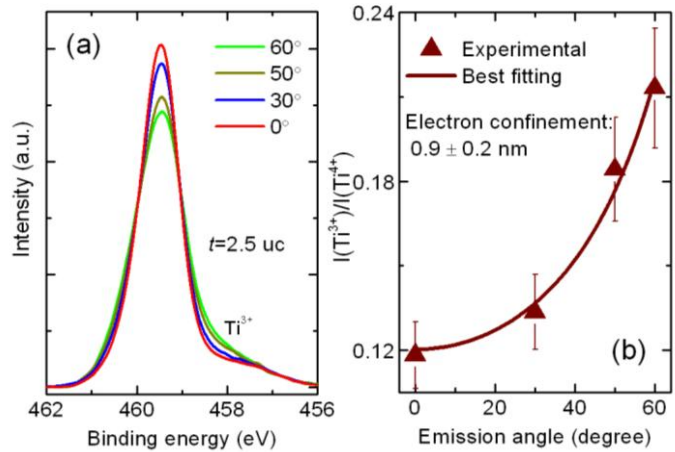

**Figure 10 Spatial confinement of the 2DEG at the GAO/STO heterointerface determined by angle-resolved XPS.** (a) The Ti $2p_{3/2}$ XPS spectra at various emission angles $\alpha$ for the $t$=2.5 uc sample; (b) The angle dependence of the ratio of Ti$^{3+}$ to Ti$^{4+}$ signal, $I(\text{Ti}^{3+})/I(\text{Ti}^{4+})$, indicates a strong confinement of the conduction layer within 0.9 nm. Error bars indicate standard deviations, ±20%, for experimental values.[36]

To determine the origin and depth-profile for the conduction in the GAO/STO heterostructures, angle resolved XPS measurements are also performed. We find that the electrons are exclusively accumulated on the otherwise empty 3$d$ shell of Ti$^{4+}$ on the STO side. The most prominent feature of the XPS data is that the Ti$^{3+}$ signal in GAO/STO heterointerfaces shows strong dependence on the photoelectrons detection angle, $\alpha$, with respect to the surface normal. An increase of the Ti$^{3+}$ signal with increasing $\alpha$, as shown in Fig. 10 (a), is clearly detected for $t$=2.5 uc with the highest Hall mobility. This further confirms that the conduction in our GAO/STO heterointerface is highly confined at the interface region. To make more quantitative analyses, we assume a simple case that the 2DEG extends from the interface to a depth, $d$, into the STO substrate.[68] The interface region is further assumed to be stoichiometric, sharp and characterized by a constant fraction, $p$, of $Ti^{3+}$ per STO unit cell. Taking into account the attenuation length of photoelectrons, the best fitting of the experimental $I(Ti^{3+})/I(Ti^{4+})$ ratios gives a $p \approx 0.31$, which equals to a $n_s \approx 2.1 \times 10^{14}$ cm$^{-2}$, and a $d$ of 0.9 nm. Therefore, the electrons at our GAO/STO heterointerface are strongly confined within approximately the first 2 uc of STO surface proximate to the interface. Note that the $n_s$ deduced here is slightly lower than that obtained from Hall data [Fig. 8(c)]. This could be due to the presence of outward diffusion of the Ti-cations into alumina films, where $Ti^{4+}$ is the dominant component.

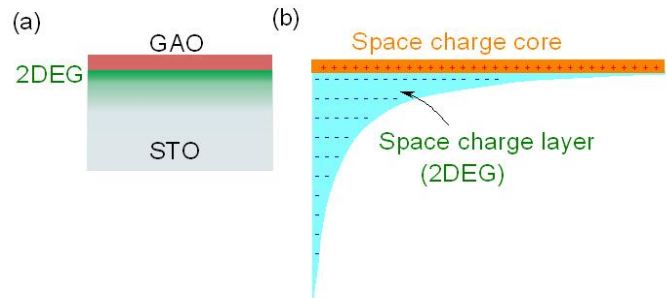

**Fig.11 Space charge layer deominated 2DEG at the GAO/STO heterointerface.**
The presence of $Ti^{3+}$ implies the formation of oxygen vacancies on the STO side. This scenario is consistent with the fact that the interfacial conductivity can be completely removed when the $Ti^{3+}$ content is significantly suppressed by annealing the sample in 1 bar pure $O_2$ at a temperature higher than 200 ºC. Such oxygen-vacancy-dominated 2DEG is expected to be formed as a consequence of chemical redox reactions occurring on the STO surface during the film growth of GAO, analogous to what has been observed in amorphous STO-based heterostructures grown at room temperature.[33] Note that the 2DEG at the crystalline GAO/STO heterointerface is formed at a high temperature of 600 ºC, where oxygen ions in STO are already highly mobile. This is normally expected to level out any difference in the depth profile of oxygen distribution in STO as discussed previously. However, this is not the case in the crystalline GAO/STO heterostructures as inferred from both Fig. 9 and Fig. 10. Moreover, the conduction at the interface of thick films, for example at $t=8$ uc, can survive the annealing at 300 ºC for 24 hours in 1 bar pure $O_2$ with only negligible changes in conductivity.[36] These features strongly suggest that the oxygen vacancies and the 2DEGs are stabilized by an interface effect, for instance the formation of a space charge region near the heterointerface as illustrated in Fig. 11.[62,63] In the scenario of space charge induced formation of 2DEGs, oxygen vacancies are enriched at the space charge core, while electrons are located in a separate space charge layer parallel to the surface. This is similar to the situation of the modulation doping observed in semiconductor 2DEGs[69] and may explain the extremely high mobility observed in the GAO/STO interface. It is worth noting that an inherent oxygen ion deficiency has been observed at the grain boundary of STO bicrystals,[70] where a considerable electron accumulation has also been predicted if the barrier height of the grain boundary is deliberately controlled.[63] The high electron mobility of STO-based oxide materials at low temperatures is generally related to the polarization shielding of the ionized defect scattering centers driven by the large dielectric constant of STO.[71] The higher mobility of our spinel/perovskite oxide interface compared to the perovskite-type oxide heterointerface may also be due to the better lattice match and thereby a more perfect structure and well-defined interface.

## 4. Conclusions and Outlook

We have reviewed our recent experimental work on 2DEGs at oxide interfaces. Despite of its early age, oxide 2DEG has already demonstrated the capability to offer great opportunities for physics and material science. Besides interface polarity, the interfacial redox reaction at oxide interfaces provides an alternative approach to explore oxide 2DEGs. Interestingly, the *a*-LAO/STO system is found to show tunable metal-insulator transitions with gated electric field,[72] analogy to what has been found in the crystalline LAO/STO interface.[15] This makes the *a*-LAO/STO system rather promising for oxide electronics, though enhancing its carrier mobility remains challenging. On the other hand, relying on interface-confined oxygen vacancies and by combining two of the largest groups of oxides, the spinel/perovskite GAO/STO interface is expected to result in plenty of new physical properties, for instance, interfacial magnetism[14] and superconductivity[13] as observed in the perovskite-type LAO/STO interface. Moreover, with a large enhancement of the electron mobility, the GAO/STO heterointerface enables the design of mesoscopic quantum devices based on complex oxide 2DEGs and opens new avenues for oxide nanoelectronics and mesoscopic physics.

**Acknowledgement**

The authors would thank the collaborations and discussions with A. J. H.M. Rijnders, G. Koster, J. E. Kleibeuker, F. Trier, D. V. Christensen, N. Bovet, N. H. Andersen, T. Kasama, W. Zhang, L. Lu, F. M. Qu, R. Giraud, J. Dufouleur, B. Buchner, T. S. Jespersen, J. Nygård, E. Stamate, S. Amoruso, J. Fleig, F. W. Poulsen and N. Bonanos.